# ON THE COMPUTATIONAL CAPABILITIES OF PHYSICAL SYSTEMS PART II: RELATIONSHIP WITH CONVENTIONAL COMPUTER SCIENCE

by David H. Wolpert

NASA Ames Research Center, N269-1, Moffett Field, CA 94035, dhw@ptolemy.arc.nasa.gov

**Abstract**: In the first of this pair of papers, it was proven that there cannot be a physical computer to which one can properly pose any and all computational tasks concerning the physical universe. It was then further proven that no physical computer C can correctly carry out all computational tasks that can be posed to C. As a particular example, this result means that no physical computer that can, for any physical system external to that computer, take the specification of that external system's state as input and then correctly predict its future state before that future state actually occurs; one cannot build a physical computer that can be assured of correctly "processing information faster than the universe does". These results do not rely on systems that are infinite, and/or non-classical, and/or obey chaotic dynamics. They also hold even if one uses an infinitely fast, infinitely dense computer, with computational powers greater than that of a Turing Machine. This generality is a direct consequence of the fact that a novel definition of computation — "physical computation" — is needed to address the issues considered in these papers, which concern real physical computers. While this novel definition does not fit into the traditional Chomsky hierarchy, the mathematical structure and impossibility results associated with it have parallels in the mathematics of the Chomsky hierarchy. This second paper of the pair presents a preliminary exploration of some of this mathematical structure. Analogues of Chomskian results concerning universal Turing Machines and the Halting theorem are derived, as are results concerning the (im)possibility of certain kinds of error-correcting codes. In addition, an analogue of algorithmic information complexity, "prediction complexity", is elaborated. A task-independent bound is derived on how much the prediction complexity of a computational task can differ for two different reference universal physical computers used to solve that task, a bound similar to the "encoding" bound governing how much the algorithm information complexity of a Turing machine calculation can differ for two reference universal Turing machines. Finally, it is proven that either the Hamiltonian of our universe proscribes a certain type of computation, or prediction complexity is unique (unlike algorithmic information complexity), in that there is one and only version of it that can be applicable throughout our universe.

**INTRODUCTION**

Recently there has been heightened interest in the relationship between physics and computation ([1-33]). This interest extends far beyond the topic of quantum computation. On the one hand, physics has been used to investigate the limits on computation imposed by operating computers in the real physical universe. Conversely, there has been speculation concerning the limits imposed on the physical universe (or at least imposed on our models of the physical universe) by the need for the universe to process information, as computers do.

To investigate this second issue one would like to know what fundamental distinctions, if any, there are between the physical universe and a physical computer. To address this issue the first of this pair of papers begins by establishing that the universe cannot contain a computer to which one can pose any arbitrary computational task. Accordingly, paper I goes on to consider computer-indexed subsets of computational tasks, where all the members of any such subset *can* be posed to the associated computer. It then proves that one cannot build a computer that can "process information faster than the universe". More precisely, it is shown that one cannot build a computer that can, for any physical system, correctly predict any aspect of that system's future state before that future state actually occurs. This is true even if the prediction problem is restricted to be from the set of computational tasks that can be posed to the computer.

This asymmetry in computational speeds constitutes a fundamental distinction between the universe and the set of all physical computers. Its existence casts an interesting light on the ideas of Fredkin, Landauer and others concerning whether the universe "is" a computer, whether there are "information-processing restrictions" on the laws of physics, etc. [10, 18]. In a certain sense, the universe is more powerful than any information-processing system constructed within it could be. This result can alternatively be viewed as a restriction on the universe as a whole — the universe cannot support the existence within it of a computer that can process information as fast as it can.

The analysis of paper I also establishes (for example) the necessarily fallible nature of retrod-iction, of control, and of observation. (This latter result can be viewed as a kind of uncertainty principle that does not rely on quantum mechanics.) The way that results of such generality are derived is by examining the underlying issues from the broad perspective of the computational character of physical systems in general, rather than that of some single precisely specified physical system. The associated mathematics does not directly involve dynamical systems like Turing machines. Rather it casts computation in terms of partitions of the space of possible worldlines of the universe. For example, to specify what input a particular physical computer has at a particular time is to specify a particular subset of all possible worldlines of the universe; different inputs to the computation correspond to different such subsets. Similar partitions specify outputs of a physical computer. Results concerning the (im)possibility of certain kinds of physical computation are derived by considering the relationship between these kinds of partitions. In its being defined in terms of such partitions, "physical computation" involves a structure that need not even be instantiated in some particular physically localized apparatus; the formal definition of a physical computer is general enough to also include more subtle non-localized dynamical processes unfolding across the entire universe.

This second paper begins with a cursory review of these partition-based definitions and results of paper I. Despite its being distinct from the mathematics of the Chomsky hierarchy, as elaborated below, the mathematics and impossibility results governing these partitions bears many parallels with that of the Chomsky hierarchy. Section 2 of this second paper explicates some of that mathematical structure, involving topics ranging from error correction to the (lack of) transitivity of computational predictability across multiple distinct computers. In particular, results are presented concerning physical computation analogues of the mathematics of Turing machines, e.g., "universal" physical computers, and Halting theorems for physical computers. In addition, an analogue of algorithmic information complexity, "prediction complexity", is elaborated. A task-independent bound is derived on how much the prediction complexity of a computational task can differ for two different reference universal physical computers used to solve that task. This bound is similar to the "encoding" bound governing how much the algorithmic information complexity of a Turing machine calculation can differ for two reference universal Turing machines. It is then proven that one of two cases must hold. One is that the Hamiltonian of our universe proscribes a certain type of computation. The other possibility is that, unlike conventional algorithmic information complexity, its physical computation analogue is unique, in that there is one and only version of it that can be applicable throughout our universe.

Throughout these papers, $\mathbf{B} \equiv \{0, 1\}$, $\mathfrak{R}$ is defined to be the set of all real numbers, '$\wedge$' is the logical *and* operator, and 'NOT' is the logical *not* operator applied to $\mathbf{B}$. To avoid proliferation of symbols, often set-delineating curly brackets will be used surrounding a single symbol, in which case that symbol is to taken to be a variable with the indicated set being the set of all values of that variable. So for example "$\{y\}$" refers to the set of all values of the variable y. In addition o(A) is the cardinality of any set A, and $2^A$ is the power set of A. $u \in U$ are the possible states of the universe, and $\hat{U}$ is the space of allowed trajectories through U. So $\hat{u} \in \hat{U}$ is a single-valued map from $t \in \mathfrak{R}$ to $u \in U$, with $u_t \equiv \hat{u}_t$ the state of the universe at time t. Note that since the universe is microscopically deterministic, $u_t$ for any t uniquely specifies $\hat{u}$. Sometimes there will be implicit constraints on $\hat{U}$. For example, we will assume in discussing any particular computer that the space $\hat{U}$ is restricted to worldlines $\hat{u}$ that contain that computer. An earlier analysis addressing some of the issues considered in this pair of papers can be found in [30].

# I. REVIEW OF DEFINITIONS AND FOUNDATIONAL RESULTS RELATED TO PHYSICAL COMPUTATION

In paper I the process by which real physical computers make predictions concerning physical systems is abstracted to produce a mathematical definition of physical computation. This section reviews that definition and the associated fundamental mathematical results. The reader is referred to paper I for more extensive discussion of the definitions.

**i) Definition of a Physical Computer**

We start by distinguishing the specification of what we want the computer to calculate from the results of that calculation:

**Definition 1:** Any *question* $q \in Q$ is a pair, consisting of a set A of *answers* and a single-valued function from $\hat{u} \in \hat{U}$ to $\alpha \in A$. $A(q)$ indicates the A-component of the pair q.

Here we restrict attention to Q that are non-empty and such that there exist at least two elements in $A(q)$ for at least one $q \in Q$. We make no other *a priori* assumptions concerning the spaces $\{A(q \in Q)\}$ and Q. In particular, we make no assumptions concerning their finiteness.

**Example 1 (conventional prediction of the future):** Say that our universe contains a system S external to our computer that is closed in the time interval [0, T], and let u be the values of the elements of a set of canonical variables describing the universe. $\alpha$ is the $t = T$ values of the components of u that concern S, measured on some finite grid G of finite precision. q is this definition of $\alpha$ with G and the like fully specified. (So q is a partition of the space of possible $u_T$, and therefore of $\hat{U}$, and $\alpha$ is an element of that partition.) Q is a set of such q's, differing in G, whose associated answers our computer can (we hope) predict correctly.

The input to the computer is implicitly reflected in its $t = 0$ physical state, as our interpretation of that state. In this example (though not necessarily in general), that input specifies what question we want answered, i.e., which q and associated T we are interested in. It also delineates one of several regions $R \subseteq \hat{U}$, each of which, intuitively, gives the $t = 0$ state of S. Throughout each such R, the system S is closed from the rest of the universe during $t \in [0, T]$. The precise R delineated further specifies a set of possible values of $u_0$ (and therefore of the Hamiltonian describing S), for example by being an element of a (perhaps irregular) finite precision grid over $U_0$, G'. If, for some R, $q(\hat{u})$ has the same value for all $\hat{u} \in R$, then this input R uniquely specifies what $\alpha$ is for any associated $\hat{u}$. If this is not the case, then the R input to the computer does not suffice to answer question q. So for any q and region R both of which can be specified in the computer's input, R must be a subset of a region $q^{-1}(\alpha)$ for some $\alpha$.

Implicit in this definition is some means for correctly getting the information R into the computer's input. In practice, this is often done by having had the computer coupled to S sometime before time 0. As an alternative, rather than specify R in the input, we could have the input contain a "pointer" telling the computer where to look to get the information R. (The analysis of these papers holds no matter how the computer gains access to R.) In addition, in practice the input, giving R, q, and T, is an element of a partition over an "input section" of our computer. In such a case, the input is itself an element of a finite precision grid over $\hat{U}$, G". So an element of G" specifies an element of G (namely q) and element of G' (namely R.)

Given its input, the computer (tries to) form its prediction for $\alpha$ by first running the laws of physics on a $u_0$ having the specified value as measured on G', according to the specified Hamiltonian, up to the specified time T. The computer then applies q(.) to the result. Finally, it writes this prediction for $\alpha$ onto its output and halts. (More precisely, using some fourth finite precision grid G''' over its output section, it "writes out" (what we interpret as) its prediction for what region in U the universe will be in at T, that prediction being formally equivalent to a prediction of a region in $\hat{U}$.) The goal is to have it do this, with the correct value of $\alpha$, by time $\tau < T$. Note that to have the computer's output be meaningful, it must specify the question q being answered as well as the answer $\alpha$, i.e., the output must be a physical state of the computer that we interpret as a question-answer pair.

Consider again the case where there is in fact a correct prediction, i.e., where R is indeed a subset of the region $q^{-1}(\alpha)$ for some $\alpha$. For this case, formally speaking, "all the computer has to do" in making its prediction is recognize which such region in the partition q that is input to the computer contains the region R that is also input to the computer. Then it must output the label of that region in q. In practice though, q and R are usually "encoded" differently, and the computer must "translate" between those encodings to recognize which region $q^{-1}(\alpha)$ contains R; this translation constitutes the "computation".

Given this definition of a question, we can now define the input and output portions of a physical computer by generalizing our example of conventional computation.

**Definition 2: i)** A (computation) *partition* is a set of disjoint subsets of $\hat{u}$ whose union equals $\hat{U}$, or equivalently a single-valued mapping from $\hat{U}$ into a non-empty space of partition-element labels. Unless stated otherwise, any partition is assumed to contain at least two elements.

**ii)** In an *output partition*, the space of partition element labels is a space of possible "outputs", {OUT}.

**iii)** In a physical computer, we require {OUT} to be the space of all pairs $\{OUT_q \in Q, OUT_\alpha \in A(OUT_q)\}$, for some Q and A(.) as defined in Def. (1). This space — and therefore the associated output partition — is implicitly a function of Q. To make this explicit, often, rather than an output partition, we will consider the full associated double (Q, OUT(.)), where OUT(.) is the output partition $\hat{u} \in \hat{U} \to OUT \in \{OUT_q \in Q, OUT_\alpha \in A(OUT_q)\}$. Also, we will find it useful to use an output partition to define an associated ("**p**rediction") partition, $OUT_p(.) : \hat{u} \to (A(OUT_q(\hat{u})), OUT_\alpha(\hat{u}))$.

**iv)** In an *input partition*, the space of partition element labels is a space of possible "inputs", {IN}.

**v)** A (*physical*) *computer* consists of an input partition and an output partition double. Unless explicitly stated otherwise, both of those partitions are required to be (separately) surjective.

Since we are restricting attention to non-empty Q, {OUT} is non-empty. We say that $OUT_q$ is the "question posed to the computer", and $OUT_\alpha$ is "the computer's answer". The surjectivity of IN(.) and OUT(.) is a restriction on {IN} and {OUT}, respectively.

While motivated in large measure by the task of predicting the future, the definition of physical computation is far broader, concerning any computation that can be cast in terms of inputs, questions about physical states of nature, and associated answers. This set of questions includes in particular any calculation that can be instantiated in a physical system in our universe, whether that question is a "prediction" or not. All such physically realizable calculations are subject to the results presented below.

Even in the context of prediction though, the definition of a physical computer presented here is much broader than computers that work by the process outlined in Ex. 1 (and therefore the associated theorems are correspondingly further-ranging in their implications). For example, the computer in Ex. 1 has the laws of physics explicitly built into its "program". But our definition allows other kinds of "programs" as well. Our definition also allows other kinds of information input to the computer besides q and a region R (which together with T constitute the inputs in Ex. 1). As discussed in paper I, we will only need to require that there be *some* t = 0 state of the computer that, by accident or by design, induces the correct prediction at t = τ. This means we do not even require that the computer's initial state IN "accurately describes" the t = 0 external universe in any meaningful sense. Our generalization of Ex. 1 preserves analogues of the grids G (in Q(.)), G" (in IN(.)) and G'" (in OUT(.)), but not of the grid G'.

In fact, our formal definition of a physical computer broadens what we mean by the "input to the computer", IN, even further. While the motivation for our definition, exemplified in Ex. 1, has the partition IN(.) "fix the initial state of the computer's inputs section", that need not be the case. IN(.) can reflect *any* attributes of $\hat{u}$. An "input" — an element of a partition of $\hat{U}$ — need not even involve the t = 0 state of the physical computer. In other words, as we use the terms here, the computer's "input" need not be specified in some t = 0 state of a physical device. Indeed, our definition does not even explicitly delineate the particular physical system within the universe that we identify with the computer. (A physical computer is simply an input partition together with an output partition.) This means we can even choose to have the entire universe "be the computer". For our purposes, we do not need tighter restrictions in our definition of a physical computer. Nonetheless, a pedagogically useful example is any localized physical device in the real world meeting our limited restrictions. No matter how that device works, it is subject to the impossibility results described below.

**ii) Intelligible computation**

Consider a "conventional" physical computer, consisting of an underlying physical system whose t = 0 state sets IN($\hat{u}$) and whose state at time $\tau$ sets OUT($\hat{u}$), as in our example above. We wish to analyze whether the physical system underlying that computer can calculate the future sufficiently quickly. In doing so, we do not want to allow any of the "computational load" of the calculation to be "hidden" in a restriction on the possible questions. Our computer possess a sufficient degree of flexibility. We impose this via the following construction (see paper I for a detailed justification):

**Definition 3:** Consider a physical computer $C \equiv (Q, IN(.), OUT(.))$ and a $\hat{U}$-partition $\pi$. A function from $\hat{U}$ into **B**, f, is an *intelligibility function* (for $\pi$) if

$$\forall \hat{u}, \hat{u}' \in \hat{U}, \pi(\hat{u}) = \pi(\hat{u}') \Rightarrow f(\hat{u}) = f(\hat{u}').$$

A set F of such intelligibility functions is an *intelligibility set* for $\pi$.

We view any intelligibility function as a question by defining A(f) to be the image of $\hat{U}$ under f. If F is an intelligibility set for $\pi$ and $F \subseteq Q$, we say that $\pi$ is *intelligible* to C with respect to F. If the intelligibility set is not specified, it is implicitly understood to be the set of all intelligibility functions for $\pi$.

We say that two physical computers $C^1$ and $C^2$ are *mutually intelligible* (with respect to the pair $\{F^i\}$) iff both $OUT^2$ is intelligible to $C^1$ with respect to $F^2$ and $OUT^1$ is intelligible to $C^2$ with respect to $F^1$.

Plugging in, $\pi$ is intelligible to C iff $\forall$ intelligibility functions f, $\exists q \in OUT_q$ such that $q = f$, i.e., such that A(q) = the image of $\hat{U}$ under f, and such that $\forall \hat{u} \in \hat{U}, q(\hat{u}) = f(\hat{u})$. Note that since $\pi$ contains at least two elements, if $\pi$ is intelligible to C, $\exists OUT_q \in \{OUT_q\}$ such that $A(OUT_q) = $ **B**, an $OUT_q$ such that $A(OUT_q) = \{0\}$, and one such that $A(OUT_q) = \{1\}$. Usually we are interested in the case where $\pi$ is an output partition of a physical computer, as in mutual intelligibility.

Intuitively, an intelligibility function for a partition $\pi$ is a mapping from the elements of $\pi$ into **B**. $\pi$ is intelligible to C if Q contains all binary-valued functions of $\pi$, i.e., if C can have posed any question concerning the universe as measured on $\pi$. This flexibility in C ensures that C's output partition isn't "rigged ahead of time" in favor of some particular question concerning $\pi$. Formally, by the surjectivity of OUT(.), the requirement of intelligibility means that $\exists \hat{u}' \in \hat{U}$ such that $\forall \hat{u} \in \hat{U}, [OUT_q(\hat{u}')](\hat{u}) = f(\hat{u})$.

### iii) Predictable computation

We can now formalize the concept of a physical computer's "making a correct prediction":

**Definition 4:** Consider a physical computer C, partition $\pi$, and intelligibility set for $\pi$, F. We say that $\pi$ is *weakly predictable* to C with respect to F iff:

i) $\pi$ is intelligible to C with respect to F, i.e., $F \subseteq OUT_q$ ;

ii) $\forall f \in F, \exists IN \in \{IN\}$ that *weakly induces* f, i.e., an IN such that:

$IN(\hat{u}) = IN$

$\Rightarrow$

$OUT_p(\hat{u}) = (A(OUT_q(\hat{u})), OUT_\alpha(\hat{u})) = (A(f), f(\hat{u}))$.

Intuitively, condition (ii) means that for all questions q in F, there is an input state such that if C is initialized to that input state, C's answer to that question q (as evaluated at $\tau$) must be correct. Note that we even allow the computer to be mistaken about what question it is answering — i.e., for $OUT_q(\hat{u})$ to not equal f — so long as C's answer is correct. We will say a computer C' with output OUT'(.) is weakly predictable to another if the partition $OUT'_p(.)$ is. If we just say "predictable" it will be assumed that we mean weak predictability.

As a formal matter, note that in the definition of predictable, even though f(.) is surjective onto A(f) (cf. Def. 3), it may be that for some IN, the set of values $f(\hat{u})$ takes on when $\hat{u}$ is restricted so that $IN(\hat{u}) = IN$ do not cover all of A(f). The reader should also bear in mind that by surjectiv-ity, $\forall$ IN $\in$ {IN}, $\exists \hat{u} \in \hat{U}$ such that IN($\hat{u}$) = IN.

### iv) Distinguishable computers

There is one final definition that we need before we can establish our unpredictability results:

**Definition 5:** Consider a set of n physical computers $\{C^i \equiv (Q^i, IN^i(.), OUT^i(.)) : i = 1, ..., n\}$. We say $\{C^i\}$ is (*input*) *distinguishable* iff $\forall$ n-tuples (IN$^1 \in \{IN^1\}$, ..., IN$^n \in \{IN^n\}$), $\exists \hat{u} \in \hat{U}$ such that $\forall$ i, IN$^i(\hat{u})$ = IN$^i$ simultaneously.

We say that $\{C^i\}$ is *pairwise* (*input*) *distinguishable* if any pair of computers from $\{C^i\}$ is distinguishable, and will sometimes say that any two such computers $C^1$ and $C^2$ "are distinguishable from each other". We will also say that $\{C^i\}$ is a *maximal* (pairwise) distinguishable set if there are no physical computers $C \notin \{C^i\}$ such that $C \cup \{C^i\}$ is a (pairwise) distinguishable set.

### iv) The impossibility of posing arbitrary questions to a computer

The first result in paper I states that for any pair of physical computers there are *always* binary-valued questions about the state of the universe that cannot even be posed to at least one of those physical computers:

**Theorem 1:** Consider any pair of physical computers $\{C^i : i = 1, 2\}$. Either $\exists$ finite intelligibility set $F^2$ for $C^2$ such that $C^2$ is not intelligible to $C^1$ with respect to $F^2$, and/or $\exists$ finite intelligibility set $F^1$ for $C^1$ such that $C^1$ is not intelligible to $C^2$ with respect to $F^1$.

Thm. 1 reflects the fact that while we do not want to have C's output partition "rigged ahead of time" in favor of some single question, we also cannot require too much flexibility of our computer. It is necessary to balance these two considerations. Accordingly, before analyzing prediction of the future, to circumvent Thm. 1 we must define a restricted kind of intelligibility set to which Thm. 1 does not apply:

**Definition 6:** An intelligibility function f for an output partition OUT(.) is *question-independent* iff $\forall \hat{u}, \hat{u}' \in \hat{U}$:

$$OUT_p(\hat{u}) = OUT_p(\hat{u}')$$
$$\Rightarrow$$
$$f(\hat{u}) = f(\hat{u}').$$

An intelligibility set as a whole is question-independent if all its elements are.

We write $C^1 > C^2$ (or equivalently $C^2 < C^1$) and say simply that $C^2$ is (weakly) *predictable* to $C^1$ (or equivalently that $C^1$ *can predict* $C^2$) if $C^2$ is weakly predictable to $C^1$ for all question-independent finite intelligibility sets for $C^2$.

Similarly, from now on we will say that $C^2$ is *intelligible* to $C^1$ without specification of an intelligibility set if $C^2$ is intelligible to $C^1$ with respect to all question-independent finite intelligibility sets for $C^2$.

Intuitively, f is question-independent if its value does not vary with q among any set of q all of which share the same A(q). As an example, say our physical computer is a conventional digital workstation. Have a certain section of the workstation's RAM be designated the "output section" of that workstation. That output section is further divided into a "question subsection" designating (i.e., "containing") a q, and an "answer subsection" designating an α. Say that for all q that can be designated by the question subsection A(q) is a single bit, i.e., we are only interested in binary-valued questions. Then for a question-independent f, the value of f can only depend on whether the answer subsection contains a 0 or a 1. It cannot vary with the contents of the question subsection.

A detailed example of a pair of mutually (question-independent) intelligible computers is presented in paper I. In addition to this explicit demonstration that Thm. 1 does not hold for question-independent intelligibility sets, examples 2, 2', and 2" of paper I establish that:

a) There are pairs of input-distinguishable physical computers, $C^1$, $C^2$, in which $C^2$ is predictable to $C^1$, $C^1 > C^2$;

b) Given $C^1$ and $C^2$ as in (a), we could have yet another computer $C^3$ that also predicts $C^2$ (i.e., such that $C^3 > C^2$) while being distinguishable from $C^1$;

c) Given $C^1$ and $C^2$ as in (a), we could have a computer $C^4$, distinguishable from both $C^1$ and $C^2$, where $C^4 > C^1$, so that $C^4 > C^1 > C^2$. We can do this either with $C^4 > C^2$ or not.

**ii) The impossibility of assuredly correct prediction**

To establish our main impossibility result in paper I we started with the following lemma:

**Lemma 1:** Consider a physical computer $C^1$. If $\exists$ any output partition $OUT^2$ that is intelligible to $C^1$, then $\exists\, q^1 \in Q^1$ such that $A(q^1) = \mathbf{B}$, a $q^1 \in Q^1$ such that $A(q^1) = \{0\}$, and a $q^1 \in Q^1$ such that $A(q^1) = \{1\}$.

This can be used to establish paper I's central theorem:

**Theorem 2**: Consider any pair of distinguishable physical computers $\{C^i : i = 1, 2\}$. It is not possible that both $C^1 > C^2$ and $C^1 < C^2$.

Restating it, Thm. 2 says that either $\exists$ finite question-independent intelligibility set for $C^1$, $F^1$, such that $C^1$ is not predictable to $C^2$ with respect to $F^1$, and/or $\exists$ finite question-independent intelligibility set for $C^2$, $F^2$, such that $C^2$ is not predictable to $C^1$ with respect to $F^2$.

Thm. 2 holds no matter how large and powerful our computers are; it even holds if the "physical system underlying" one or both of our computers is the whole universe. It also holds if instead $C^2$ is the rest of the physical universe external to $C^1$. A set of implications of Thm. 2 for various kinds of physical prediction scenarios are discussed in paper I. As also discussed there, impossi-bility results that are in some senses even stronger than those associated with Thm. 2 hold when we do not restrict ourselves to distinguishable computers, as we do in Thm. 2.

## 3. THE MATHEMATICAL STRUCTURE RELATING PHYSICAL COMPUTERS

There is a rich mathematical structure governing the possible predictability relationships among sets of physical computers, especially if one relaxes the presumption (obtaining in much of paper I) that the universe can contain multiple copies of C. This section presents some of that structure.

**i) The graphical structure over a set of computers induced by weak predictability**

While it directly concerns pairs of physical computers, Thm. 2 also has implications for the predictability relationships within sets of more than two computers. An example is the following:

**Corollary 1:** It is not possible to have a fully distinguishable set of n physical computers $\{C^i\}$ such that $C^1 > C^2 > ... > C^n > C^1$.

**Proof:** Hypothesize that the corollary is wrong. Define the composite device $C^* \equiv (IN^*(.) \equiv \Pi_{i=1}^{n-1} IN^i(.), Q^1, OUT^1(.))$. Since $\{C^i\}$ is fully distinguishable, $IN^*(.)$ is surjective. Therefore $C^*$ is a physical computer.

Since by hypothesis $C^n$ is intelligible to $C^{n-1}$, $\exists\ OUT^{n-1}{}_q$ such that $A(OUT^{n-1}{}_q) = \mathbf{B}$. Also, since $C^{n-2} > C^{n-1}$, $\exists\ IN^{n-2} \in \{IN^{n-2}\}$ such that $\forall\ \hat{u} \in \hat{U}$ for which $A(OUT^{n-1}{}_q(\hat{u})) = \mathbf{B}$, $IN^{n-2}(\hat{u}) = IN^{n-2} \Rightarrow OUT^{n-2}{}_\alpha(\hat{u}) = OUT^{n-1}{}_\alpha(\hat{u})$. Iterating and exploiting full distinguishability, $\exists\ (IN^1, ..., IN^{n-2})$ such that $\forall\ \hat{u} \in \hat{U}$ for which $A(OUT^{n-1}{}_q(\hat{u})) = \mathbf{B}$, $(IN^1(\hat{u}), .., IN^{n-2}(\hat{u})) = (IN^1, ..., IN^{n-2}) \Rightarrow OUT^*(\hat{u}) = OUT^1(\hat{u}) = OUT^{n-1}(\hat{u})$. The same holds when we restrict $\hat{u}$ so that the space $A(OUT^{n-1}{}_q(\hat{u})) = \{1\}$, and when we restrict $\hat{u}$ so that $A(OUT^{n-1}{}_q(\hat{u})) = \{0\}$.

Since by hypothesis $C^n$ is intelligible to $C^{n-1}$, and since $IN^*(.)$ is surjective, this result means that $C^n$ is predictable to $C^*$. Conversely, since $C^n > C^1$ by hypothesis, the output partition of $C^*$ is predictable to $C^n$, and therefore $C^*$ is. Finally, since $\{C^i\}$ is fully distinguishable, $C^*$ and $C^n$ are distinguishable. Therefore Thm. 2 applies, and by using our hypothesis we arrive at a contradiction. **QED.**

What are the general conditions under which two computers can be predictable to one another? By Thm. 2, we know they aren't if they're input-distinguishable. What about if they're one and the same? No physical computer is input-distinguishable from itself, so Thm. 2 doesn't apply to this issue. However it still turns out that Thm. 2's implication holds for this issue:

**Theorem 3:** No physical computer is predictable to itself.

**Proof.** Assume our corollary is wrong, and some computer C is predictable to itself. Since by definition predictability implies intelligibility, we can apply Lemma 1 to establish that there is a $q \in OUT_q$, q', such that $A(q') = \mathbf{B}$. Therefore one question-independent intelligibility function for C is the function f from $\hat{u} \in \hat{U} \to \mathbf{B}$ that equals 1 if $A(OUT_q(\hat{u})) = \mathbf{B}$ and $OUT_\alpha(\hat{u}) = 0$, and equals 0 otherwise. Therefore by hypothesis $\exists$ IN $\in \{IN\}$ such that $IN(\hat{u}) = IN \Rightarrow A(OUT_q(\hat{u})) = \mathbf{B}$ and $OUT_\alpha(\hat{u}) = f(\hat{u})$. But if $A(OUT_q(\hat{u})) = \mathbf{B}$, then $f(\hat{u}) = NOT[OUT_\alpha(\hat{u})]$, by definition of f(.). Since IN is surjective, this means that there is at least one $\hat{u} \in \hat{U}$ such that $A(OUT_q(\hat{u})) = \mathbf{B}$ and $OUT_\alpha(\hat{u}) = NOT[OUT_\alpha(\hat{u})]$. This is impossible. **QED.**

Intuitively, this result holds due to the fact that a computer cannot make as its prediction the logical inverse of its prediction. An important corollary of this result is that no output partition is predictable to a physical computer that has that output partition. Combining Thm. 3 and Coroll. 1 and identifying the predictability relationship with an edge in a graph, we see that fully distinguishable sets of physical computers constitute (unions of) directed acyclic graphs.

**ii) Weak predictability and variants of error correction**

When considering sets of more than two computers, it is important to realize that while it is symmetric, the input-distinguishability relation need not be transitive. Accordingly, separate pairwise distinguishable sets of computers may partially "overlap" one another. Similarly, stipulating the values of the inputs of any two computers in a pairwise-distinguishable set may force some of the other computers in that set to have a particular input value.

Coroll. 1 does not apply to such a set. As it turns out though, Thm. 2 still has strong implications even for a set of more than two computers that is not fully distinguishable, so long as the set is pairwise distinguishable. Define a *god computer* as any physical computer in a pairwise distinguishable set such that all other physical computers in that set are predictable to the god computer. Then by Thm. 2, each such set can contain at most one god computer. There is at most one computer in any pairwise distinguishable set that can correctly predict the future of all other members of that set, and more generally at most one that can accurately predict the past of, observe, and/or control any system in that set (see paper I). In particular, for any human being physical computer, for any pairwise distinguishable set of computers including that human, there can be at most one god computer. (Lest one read too much into the phrase "god computer", note that like any other computer, a god computer is merely a set of partitions, and need not correspond to any localized physical apparatus.)

Even a god computer may not be able to correctly predict all other computers in its distinguishable set simultaneously. The input value it needs to adopt to correctly predict some $C^2$ may preclude it from correctly predicting some $C^3$ and vice-versa. One way to analyze this issue is to consider a composite partition $OUT^{2\times 3}$ defined by the output partitions of $C^2$ and $C^3$. We can then investigate whether and when our god computer can weakly predict the composite output partition. The following definition formalizes this:

**Definition 7:** Consider a pairwise distinguishable set $\{C^i\}$ with god computer $C^1$. Define the par-titions $OUT^{i \times j}(\hat{u} \in \hat{U}) \equiv (OUT^{i \times j}_q(\hat{u}), OUT^{i \times j}_\alpha(\hat{u}))$, where each answer map $OUT^{i \times j}_\alpha(\hat{u}) \equiv (OUT^1_\alpha(\hat{u}), OUT^2_\alpha(\hat{u}))$, and each question $[OUT^{i \times j}_q(\hat{u})] \equiv$ the mapping given by $\hat{u}' \in \hat{U} \to ([OUT^1_q(\hat{u})](\hat{u}'), [OUT^2_q(\hat{u})](\hat{u}'))$. Then $C^1$ is *omniscient* if $OUT^{2 \times 3 \times \cdots}$ is weakly predictable to $C^1$.

Intuitively, $OUT^{i \times j}$ is just the double partition $(OUT^i(.), OUT^j(.)) = ((OUT^i_q(.), OUT^i_\alpha(.)), (OUT^j_q(.), OUT^j_\alpha(.)))$, re-expressed to be in terms of a single question-valued partition and a single answer-valued partition. To motivate this re-expression, for any two questions $q^i \in Q^i$ and $q^j \in Q^j$, let $q^i \times q^j$ be the ordered product of the partitions $q^i$ and $q^j$; it is the partition assigning to every point $\hat{u}' \in \hat{U}$ the label $(q^i(\hat{u}'), q^j(\hat{u}'))$. Then if $OUT^i_q(\hat{u})$ is the question $q^i$ and $OUT^j_q(\hat{u})$ is the question $q^j$, $OUT^{i \times j}_q(\hat{u})$ is the question $q^i \times q^j$. $OUT^{i \times j}_\alpha$ is defined similarly, only with one fewer levels of "indirection", since answer components of output partitions are not themselves partitions (unlike question components).

Note that even though any $OUT^i(.)$ and $OUT^j(.)$ are both surjective mappings, $OUT^{i \times j}$ need not be surjective onto the set of quadruples $\{q^i \in Q^i, q^j \in Q^j, \alpha^i \in A(Q^i), \alpha^j \in A(Q^j)\}$. It is straight-forward to verify that an omniscient computer is a god computer.

In general, one might presume that two non-god computers in a pairwise-distinguishable set could have the property that, while individually they cannot predict everything, considered jointly they would constitute a god computer, if only they could work cooperatively. An example of such cooperativity would be having one of the computers predict when the other one's prediction is wrong. It turns out though that under some circumstances the mere presence of some other computer in that pairwise distinguishable set may make such error-correction impossible, if that other computer is omniscient.

As an example of this, say we have three pair-wise distinguishable computers $C^1, C^2, C^3$, where $C^3$ always answers with a bit (i.e., $\nexists\, q^3 \in Q^3$ such that $A(q^3) \not\subseteq \mathbf{B}$). We will want $C^2$'s output to "correct" $C^3$'s predictions, and have those predictions potentially concern $C^1$. So have $C^1$ be intelligible to $C^3$. As a technical condition, assume not only that $C^3$'s output can be any of its possible question-answer pairs, but also that for any of its questions, for any of the associated possible answers, there are situations where that answer is correct (so that $C^2$ should leave $C^3$'s answer alone in those situations). Then it turns out that due to Thm. 2, if $C^1$ is omniscient, it is not possible that $C^2$ always correctly outputs a bit saying whether $C^3$'s answer is the correct response to $C^3$'s question. More formally,

**Corollary 2:** Consider three pair-wise distinguishable computers $C^1$, $C^2$, $C^3$, where $\nexists \; q^3 \in Q^3$ such that $A(q^3) \not\subseteq \mathbf{B}$. Assume that $C^1$ is an omniscient computer, and that $C^1$ is intelligible to $C^3$. Finally, assume that $\forall$ pairs $(q^3 \in Q^3, \alpha^3 \in A(q^3))$, $\exists \; \hat{u} \in \hat{U}$ such that both $OUT^3_q(\hat{u}) = q^3$ and $q^3(\hat{u}) = \alpha^3$ (i.e., $[OUT^3_q(\hat{u})](\hat{u}) = \alpha^3$). Then it is not possible that $\forall \; \hat{u} \in \hat{U}$, $OUT^2_\alpha(\hat{u}) = 1$ if $[OUT^3_q(\hat{u})](\hat{u}) = OUT^3_\alpha(\hat{u})$, 0 otherwise.

**Proof:** Hypothesize that the corollary is wrong. Construct a composite device $C^{2\text{-}3}$, starting by having $IN^{2\text{-}3}(.) \equiv OUT^3_q(.)$, $Q^{2\text{-}3} = Q^3$ and $OUT^{2\text{-}3}_q(.) = OUT^3_q(.)$. Next define the question $\theta$ by the rule $\theta(\hat{u}) \equiv NOT[OUT^3_\alpha(\hat{u})]$ if $OUT^2_\alpha(\hat{u}) = 0$, $\theta(\hat{u}) \equiv OUT^3_\alpha(\hat{u})$ otherwise. (N.b. no assumption is made that $\theta \in Q^{2\text{-}3}$.) To complete the definition of the composite computer $C^{2\text{-}3}$, have $OUT^{2\text{-}3}_\alpha(\hat{u}) = \theta(\hat{u})$.

Now by our hypothesis, $\forall \; \hat{u} \in \hat{U}$, $\theta(\hat{u}) = [OUT^3_q(\hat{u})](\hat{u})$. By the last of the conditions specified in the corollary, this means that $\forall \; (q^{2\text{-}3} \in Q^{2\text{-}3}, \alpha^{2\text{-}3} \in A(q^{2\text{-}3}))$, $\exists \; \hat{u}$ such that $OUT^{2\text{-}3}_q(\hat{u}) = q^{2\text{-}3}$ and $OUT^{2\text{-}3}_\alpha(\hat{u}) = \alpha^{2\text{-}3}$. So $C^{2\text{-}3}$ allows all possible values of $\{OUT^{2\text{-}3}\}$, as a physical computer must. Due to surjectivity of $OUT^3_q$, it also allows all possible values of the space $\{IN^{2\text{-}3}\}$. To complete the proof that $C^{2\text{-}3}$ is a (surjective) physical computer, we must establish that $OUT^{2\text{-}3}_\alpha(\hat{u}) \in A(OUT^{2\text{-}3}_q(\hat{u})) \; \forall \; \hat{u} \in \hat{U}$. To do this note that if for example $A(OUT^{2\text{-}3}_q(\hat{u})) = A(OUT^3_q(\hat{u})) = \{1\}$, then since it is always the case that the $OUT^{2\text{-}3}_\alpha(\hat{u}) = [OUT^{2\text{-}3}_q(\hat{u})](\hat{u}) = [OUT^3_q(\hat{u})](\hat{u})$, $OUT^{2\text{-}3}_\alpha(\hat{u}) = 1$. Similarly $OUT^{2\text{-}3}_\alpha(\hat{u}) \in A(OUT^{2\text{-}3}_q(\hat{u}))$ when $A(OUT^{2\text{-}3}_q(\hat{u})) = \{0\}$. Finally, if $A(OUT^{2\text{-}3}_q(\hat{u})) = \mathbf{B}$, then the simple fact that $OUT^{2\text{-}3}_\alpha(\hat{u}) \in \mathbf{B}$ always means that $OUT^{2\text{-}3}_\alpha(\hat{u}) \in A(OUT^{2\text{-}3}_q(\hat{u}))$.

Since $C^1$ is intelligible to $C^3$ and $Q^{2-3} = Q^3$, $C^1$ is intelligible to $C^{2-3}$. Moreover, given any question $q^{2-3} \in Q^{2-3}$, $\exists$ associated $IN^{2-3} \in \{IN^{2-3}\}$ such that $\forall \hat{u} \in \hat{U}$ for which $IN^{2-3}(\hat{u}) = IN^{2-3}$, $OUT^{2-3}(\hat{u}) = q^{2-3}$. But as was just shown, $OUT^{2-3}_\alpha(\hat{u}) = q^{2-3}(\hat{u})$ for that $\hat{u}$. Therefore $C^1$ is predictable to $C^{2-3}$.

Next, since $C^1$ is omniscient, $OUT^{2\times 3}$ is intelligible to $C^1$. Therefore any binary function of the regions defined by quadruples $(A(OUT^2_q(\hat{u})), A(OUT^3_q(\hat{u})), OUT^2_\alpha(\hat{u}), OUT^3_\alpha(\hat{u}))$ is an element of $Q^1$. Any single such region is wholly contained in one region defined by the pair $(A(OUT^{2-3}_q(\hat{u})), OUT^{2-3}_\alpha(\hat{u}))$ though. Therefore any binary function of the regions defined by such pairs is an element of $Q^1$. Therefore $C^{2-3}$ is intelligible to $Q^1$. Similarly, the value of any such binary function must be given by $OUT^1_\alpha(\hat{u})$ whenever $IN^1(\hat{u})$ equals some associated $IN^1$. So $C^{2-3}$ is predictable to $C^1$.

Finally, since $C^1$ and $C^3$ are input-distinguishable, so are $C^1$ and $C^{2-3}$, and therefore Thm. 2 applies. This establishes that our hypothesis results in a contradiction. **QED.**

This result even holds if $OUT^{2\times 3}$ is only intelligible to $C^1$, without necessarily being predictable to it.

Coroll. 2 can be viewed as a restriction on the efficacy of any error correction scheme in the presence of a (distinguishable) omniscient computer. There are other restrictions that hold even in the absence of such a third computer. An example is the following implication of Thm. 2:

**Corollary 3:** Consider two distinguishable mutually intelligible physical computers $C^1$ and $C^2$, where both $A(OUT^1_q) \subseteq \mathbf{B}$ and $A(OUT^2_q) \subseteq \mathbf{B}$ $\forall$ $OUT^1_q \in Q^1$ and $OUT^2_q \in Q^2$. It is impossible that $C^1$ and $C^2$ are "anti-predictable" to each other, in the sense that for each of them, the prediction they make concerning the state of the other can always be made to be wrong by appropriate choice of input.

**Proof:** By assumption $C^1$ and $C^2$ are mutually intelligible. So what we must establish is whether for both of them, for all intelligibility functions concerning the other one, there exists an appropriate value of $IN^i$ such that that intelligibility function is incorrectly predicted.

Hypothesize that the corollary is wrong. Then $\forall$ question-independent intelligibility functions for $C^1$, $f^1$, $\exists\ IN^2 \in \{IN^2\}$ such that $IN^2(\hat{u}) = IN^2$ implies that $[A(OUT^2_q(\hat{u})) = NOT[A(f^1)]] \wedge [OUT^2_\alpha(\hat{u}) = NOT[f^1(\hat{u})]]$. However by definition of question-independent intelligibility functions, given any such $f^1$, there must be another question-independent intelligibility function for $C^1$, $f^3$, defined by $f^3(.) \equiv NOT(f^1(.))$. Therefore $\exists\ IN^2 \in \{IN^2\}$ such that $IN^2(\hat{u}) = IN^2$ implies that $[A(OUT^2_q(\hat{u})) = A(f^3)] \wedge [OUT^2_\alpha(\hat{u}) = f^3(\hat{u})]$.

This NOT(.) transformation bijectively maps the set of all question-independent intelligibility functions for $C^2$ onto itself. Since that set is finite, this means that the image of the set under the NOT(.) transformation is the set itself. Therefore our hypothesis means that all question-independent functions for $C^1$ can be predicted correctly by $C^2$ for appropriate choice of $IN^2 \in \{IN^2\}$. By similar reasoning, we see that $C^1$ can always predict $C^2$ correctly. Since $C^1$ and $C^2$ are distinguishable, we can now apply Thm. 2 and arrive at a contradiction. **QED.**

### iii) Strong predictability

At the other end of the spectrum from distinguishable computers is the case where one computer's input can fix another's, either by being observed by that other computer or by setting that other computer's input more directly. The following variant of predictability captures this relationship:

**Definition 8:** Consider a pair of physical computers $C^1$ and $C^2$. We say that $C^2$ is *strongly predictable* to $C^1$ (or equivalently that $C^1$ *can strongly predict* $C^2$), and write $C^1 >> C^2$ (or equivalently $C^2 << C^1$) iff:

  i) $C^2$ is intelligible to $C^1$;

  ii) $\forall$ question-independent intelligibility functions for $C^2$, $q^1$, $\forall\ IN^2 \in \{IN^2\}$,

   $\exists\ IN^1 \in \{IN^1\}$ that *strongly induces* the pair $(q^1, IN^2)$, i.e., such that:

$$IN^1(\hat{u}) = IN^1$$
$$\Rightarrow$$
$$[OUT^1_p(\hat{u}) = (A(q^1), q^1(\hat{u}))] \wedge [IN^2(\hat{u}) = IN^2].$$

Intuitively, if $C^1$ can strongly predict $C^2$, then for any $IN^2$ and associated implication $OUT^2_p$ — for any computation $C^2$ might undertake — there is an input to $C^1$ that is uniquely associated with $IN^2$ and that causes $C^1$ to output (any desired question-independent intelligibility function of) $OUT^2_p$. Intuitively, there is some invertible "translating" map that takes $C^2$'s input and "encodes" it in $C^1$'s input, in such a way that $C^1$ can "emulate" $C^2$ running on $C^2$'s input, and thereby produce $C^2$'s associated output. In this way $C^1$ can emulate $C^2$, much like universal Turing machines can emulate other Turing machines. (Recall the definition of universal Turing machine, and see the definition of a universal physical computer below.)

Strong predictability of a computer implies weak predictability of that computer. (Unlike with weak predictability, there is no such thing as strong predictability of a partition.) So for example both Thm. 3 and Coroll. 1 still hold if they are changed by replacing weak predictability with strong predictability. However weak predictability does not imply strong predictability. Moreover, the mathematics for sets of physical computers some of which are strongly predictable to each other (and therefore not distinguishable) differs in some respects from that when all the computers are distinguishable (the usual context for investigations of weak predictability). An example is the following result, which shows that strong predictability always is transitive, unlike weak predictability (cf. Ex. 2" in paper I):

**Theorem 4:** Consider three physical computers $\{C^1, C^2, C^3\}$, and a partition $\pi$, where both $C^3$ and $\pi$ are intelligible to $C^1$.

  i) $C^1 \gg C^2 > \pi \Rightarrow C^1 > \pi$;
  ii) $C^1 \gg C^2 \gg C^3 \Rightarrow C^1 \gg C^3$.

**Proof:** To prove (i), let f be any question-independent intelligibility function for $\pi$. By Lemma 1, the everywhere 0-valued question-independent intelligibility function of $\pi$ is contained in $Q^1$, and since $C^1 > C^2$, there must be an $IN^1$ such that $IN^1(\hat{u}) = IN^1 \Rightarrow OUT^1_\alpha(\hat{u}) = 0$. The same is true for the everywhere 1-valued function. Therefore to prove the claim we need only establish that for every question-independent intelligibility function for $\pi$, f, for which $A(f) = \mathbf{B}$, $f \in Q^1$, and there exists an $IN^1$ such that $IN^1(\hat{u}) = IN^1 \Rightarrow OUT^1_\alpha(\hat{u}) = f(\hat{u})$. Restrict attention to such f from now on.

Define a question-independent intelligibility function for $C^2$, $I^2$, such that $A(I^2) = \mathbf{B}$, and such that for all $\hat{u}$ for which $A(OUT_q(\hat{u})) = \mathbf{B}$, $I^2(\hat{u}) = OUT^2_\alpha(\hat{u})$. (Note that since $C^2 > \pi$, there both exist $\hat{u}$ for which $OUT^2_p(\hat{u}) = (\mathbf{B}, 1)$ and $\hat{u}$ such that $OUT^2_p(\hat{u}) = (\mathbf{B}, 0)$.) Now by hypothesis, for any of the f we are considering, $\exists IN^2_f \in \{IN^2\}$ such that $IN^2(\hat{u}) = IN^2_f \Rightarrow OUT^2_p(\hat{u}) = (\mathbf{B}, f(\hat{u}))$. However the fact that $C^1 \gg C^2 \Rightarrow \exists IN^1 \in \{IN^1\}$ such that $IN^1(\hat{u}) = IN^1 \Rightarrow IN^2(\hat{u}) = IN^2_f$ and such that $OUT^1_p(\hat{u}) = (A(I^2), I^2(\hat{u})) = (\mathbf{B}, I^2(\hat{u}))$. Since $IN^2(\hat{u}) = IN^2_f$ for such a $\hat{u}$, $A(OUT^2_\alpha(\hat{u})) = \mathbf{B}$, and therefore $I^2(\hat{u}) = OUT^2_\alpha(\hat{u})$. So $OUT^2_p(\hat{u})$ for such a $\hat{u}$ equals $(\mathbf{B}, OUT^2_\alpha(\hat{u}))$. So for that $IN^1$, $OUT^1_p(\hat{u}) = (A(f), f(\hat{u}))$.

This establishes (i). The proof for (ii) goes similarly, with the redefinition that $IN^1_f$ fixes the value of $IN^3$ as well as ensuring that $OUT^2_p(\hat{u}) = (A(f), f(\hat{u}))$. **QED.**

Strong predictability obeys the following result which is analogous to both Thm.'s 2 and 3:

**Theorem 5:** Consider any pair of physical computers $\{C^i: i = 1, 2\}$. It is not possible that both $C^1 \gg C^2$ and $C^1 \ll C^2$.

**Proof:** Choose any $IN^2$. For any question-independent intelligibility function of $OUT^2_p$, f, there must exist an $IN^1_f \in \{IN^1\}$ that strongly induces $IN^2$ and f, since $C^1 \gg C^2$. Label any such $IN^1$ as $IN^1_f$ ($IN^2$ being implicitly fixed). So for any such f, $\{\hat{u} : IN^1(\hat{u}) = IN^1_f\} \subseteq \{\hat{u} : IN^2(\hat{u}) = IN^2\}$. However since $OUT^2_p$ is not empty, there are at least two question-independent intelligibility functions of $OUT^2_p$, $f_1$ and $f_2$, where $A(f_1) \neq A(f_2)$ (cf. Lemma 1). Moreover, the intersection $\{\hat{u} : IN^1(\hat{u}) = IN^1_{f_1}\} \cap \{\hat{u} : IN^1(\hat{u}) = IN^1_{f_2}\} = \emptyset$, since these two sets induce different $A(OUT^1_q)$ (namely $A(f_1)$ and $A(f_2)$, respectively). This means that $\{\hat{u} : IN^1(\hat{u}) = IN^1_{f_1}\} \subset \{\hat{u} : IN^2(\hat{u}) = IN^2\}$. On the other hand, for the same reasons, there must also exist an $IN^2$ that strongly induces $IN^1_{f_1}$. Therefore $\exists\ IN^{2'}$ such that $\{\hat{u} : IN^2(\hat{u}) = IN^{2'}\} \subset \{\hat{u} : IN^1(\hat{u}) = IN^1_{f_1}\}$. So $\{\hat{u} : IN^2(\hat{u}) = IN^{2'}\} \subset \{\hat{u} : IN^2(\hat{u}) = IN^2\}$. This is not compatible with the fact that $IN^2(.)$ is a partition. **QED.**

Many of the conditions in the preceding results can be weakened and the associated conclusions still hold. Indeed, this is even true for Thm. 2, where we can weaken the definition of "intelligibility" and still establish the impossibility of having both $C^1 > C^2$ and $C^2 > C^1$. (For example, that impossibility will still obtain even if neither $C^1$ nor $C^2$ contains **B**-valued questions, if they instead contain all possible functions mapping each others' values of $OUT_p$ onto $\{0, 1, 2\}$.) These weakened version are usually more obscure though, which is why they are not presented here.

**iv) Physical computation analogues of Halting theorems in Turing machine theory**

There are several ways that one can relate the mathematical structure of physical computation to that of conventional computer science. Here we sketch the salient concepts for one such relation coupling physical computation and the mathematical structure governing Turing machines (TMs).

A TM is a device that takes in an input string on an input tape, then based on it produces a sequence of output strings, either "halting" at some time with a final output string, or never halting. If desired, the fact that the halt state has / hasn't been entered by any time can be reflected in a special associated pattern in the output string, in which case the sequence of output strings can always be taken to be infinite. As explicated above, in the real world inputs and (sequences of) outputs are elements of partitions of $\hat{U}$. So in one translation of TMs to physical computers, strings on tapes are replaced with elements of the partitions $IN(.)$ and $OUT(.)$. Rather than through a set of internal states, read/write operations, state-transition rules, etc., the transformation of inputs to outputs in a physical computer is achieved simply through the definition of the pair of an associated input partition and output partition. For a TM that declares in its output string whether it has halted, the physical computation analogue of whether a computation will ever halt is simply whether $\hat{u}$ is in some special subset of {OUT}. Although not formally required, in the real world IN(.) and OUT(.) usually differ. In this they are analogous to TM's with multiple tapes rather than conventional single-tape TMs.

An alternative to identifying the full output partition of a physical computer with a TM's output tape, motivated by the definition of predictability, is to identify the coarser partition $\hat{u} \rightarrow OUT_p(\hat{u})$ with a TM's output tape. (This is loosely analogous to a TM's being able to overwrite the "question" originally posed on its tape when producing its "answer" on that tape.) We will adopt this identification from now on, and use it to identify the physical computation analogue of a TM as an input partition together with the surjective mapping $\hat{u} \rightarrow OUT_p(\hat{u})$ of an associated output partition.

This identification motivates several analogues of the Halting theorem. Since whether a particular physical computer $C^2$ "halts" or not can be translated into whether its output is in a particular region, the question of whether $C^2$ halts is a particular intelligibility function of $C^2$. Correctly answering the question of whether $C^2$ halts means predicting that intelligibility function of $C^2$. In the context of physical computation it is natural to broaden the issue to concern all intelligibility functions of $C^2$. Accordingly, in this analogue of the claim resolved for TM's (in the negative) by the Halting theorem, one asks if it is possible to construct a physical computer $C^1$ that can predict any computer $C^2$. To answer this, consider the case where $C^2$ is a copy of $C^1$ (cf. Def. 2(v) of paper I for a formal definition of a physical computer's "copy"). Then by applying Thm.'s 2, 3 and 5, one sees that the answer is no, in agreement with the Halting theorem. (See also Coroll. 3.)

There exist a number of alternative physical computer analogues of the Halting problem. Though not pursued at length here, it is worth briefly presenting one such alternative. This alternative is motivated by arguing that, in the real world, one is not interested so much in whether the computation will ever "halt", but rather whether the associated output is "correct". If we take "correct" to be relative to a particular question, this motivates the following alternative analogue of the Halting theorem:

**Theorem 6:** Given a set of physical computers $\{C^i\}$, $\nexists\ C^1 \in \{C^i\}$ such that $\forall\ C^2 \in \{C^i\}$,

i) $C^2$ is intelligible to $C^1$;

ii) $\forall\ q^2 \in Q^2$, $\exists\ IN^1 \in \{IN^1\}$ such that $IN^1(\hat{u}) = IN^1 \Rightarrow OUT^1_\alpha(\hat{u}) = 1$ iff $q^2(\hat{u}) = OUT^2_\alpha(\hat{u})$.

**Proof:** Choose $C^2$ such that $OUT^2(.) = OUT^1(.)$. (If need be, to do this simply choose $C^2 = C^1$.) Then in particular, $OUT^1_\alpha(.) = OUT^2_\alpha(.)$. Now since $C^2$ is intelligible to $C^1$ by hypothesis, by Lemma 1 $\exists\ q^1 \in Q^1$ such that $A(q^1) = \{0\}$, and therefore $\exists\ q^2 \in Q^2$ such that $A(q^2) = \{0\}$. For that $q^2$, $OUT^1_\alpha(\hat{u}) = 1$ iff $0 = OUT^1_\alpha(\hat{u})$, which is impossible. **QED.**

A TM $T^1$ can emulate a TM $T^2$ if for any input for $T^2$, $T^1$ produces the same output as $T^2$ when given an appropriately modified version of that input. (Typically, the "modification" involves pre-pending an encoding of $T^2$ to that input.) The analogous concept for a physical computer is strong predictability; o ne physical computer can "emulate" another if it can strongly predict that other one. Intuitively, the two components of $T^1$'s emulating $T^2$, involving $T^2$'s input and its computational behavior, respectively, correspond to the two components of the requirement concerning $IN^1$ values that occur in the definition of strong predictability. The requirement concerning $IN^1$ values that is imposed by ensuring that $OUT^1_p(\hat{u}) = (A(q), q(\hat{u}))$ for any q (that is an intelligibility function) for $C^2$ is analogous to encoding (the computational behavior of) the TM $T^2$ in a string provided to the emulating TM, $T^1$. Requiring as well that the value $IN^1$ ensures that $IN^2(\hat{u}) = IN^2$ is analogous to also including an "appropriately modified" version of $T^2$'s input in the string provided to $T^1$. (Note that any mapping taking $IN^2 \in \{IN^2\}$ to an $IN^1$ that in turn induces that starting $IN^2$ is invertible, by construction.) This motivates the following defini-tion of the analogue of a universal TM:

**Definition 9:** A *universal* physical computer for a set of physical computers is a member of that set that can strongly predict all other members of that set.

Note that rather than reproduce the output of a computer it is strongly predicting, a universal physical computer produces the value of an intelligibility function applied to that output. This allows the computers in our set to have different output spaces from the universal physical computer. However it contrasts with the situation with conventional TM's, being a generalization of such TM's.

v) **Prediction complexity**

In computer science theory, given a universal TM T, the algorithmic complexity of an output string s is defined as the length of the smallest input string s' that when input to T produces s as output. To construct our physical computation analogue of this, we need to define the "length" of an input region of a physical computer. To do this we start with the following pair of definitions:

**Definition 10:** For any physical computer C with input space $\{IN\}$:

**i)** Given any partition $\pi$, a *(weak) prediction input set* (of C, for $\pi$) is any set $s \subseteq \{IN\}$ such that both every intelligibility function for $\pi$ is weakly induced by an element of s, and for any proper subset of s at least one such function is not weakly induced. We write the space of all weak prediction input sets of C for $\pi$ as $C^{-1}(\pi)$.

**ii)** Given any other physical computer C' with input space $\{IN'\}$ for which the set of all question-independent intelligibility functions is $\{f'\}$, a *strong prediction input set* of C, for the triple C', $in' \subseteq \{IN'\}$, and $f' \subseteq \{f'\}$, is any set $s \subseteq \{IN\}$ such that both every pair (f' $\in f'$, IN' $\in in'$) is strongly induced by a member of s, and for any proper subset of s at least one such pair is not strongly induced. We write the space of all strong prediction input sets (of C, for C', $in'$, and $f'$) as

$C^{-1}(C', in', f')$.

Intuitively, the prediction set of C for $\pi$ / C' is a minimal subset of {IN} that is needed by C for $\pi$ / C' to be predictable to C. In the case of strong prediction, we provide the associated definition the extra flexibility of being able to restrict what intelligibility functions are being considered.

Now, to define the physical computation analogue of algorithmic information complexity, identify the "length of an input string" with the negative logarithm of the volume of a subset of the partition IN(.):

**Definition 11:** Given a physical computer C and a measure $d\mu$ over $\hat{U}$:

**i)** Define V($in \subseteq$ {IN}) as the measure of the set of all $\hat{u} \in \hat{U}$ such that IN($\hat{u}$) $\in in$, and define the *length* of *in* (with respect to IN(.)) as $\mathbf{l}(in) \equiv -\ln[V(in)]$;

**ii)** Given a partition $\pi$ that is predictable to a physical computer C, define the *prediction complexity* of $\pi$ (with respect to C), $\mathbf{c}(\pi \mid C)$, as $\min_{\rho \in C^{-1}(\pi)} [\mathbf{l}(\rho)]$.

We are primarily interested in prediction complexities of binary partitions, in particular of the binary partitions induced by the separate single elements of multi-element partitions. (The binary partition induced by some $p \in \pi'$ is $\{\hat{u}$ s.t. $\pi'(\hat{u}) = p$, $\hat{u}$ s.t. $\pi'(\hat{u}) \neq p\}$.) To see what Def. 11(ii) means for such a partition, say you are given some set $\sigma \subset \hat{U}$ (i.e., you are given a binary partition of $\hat{U}$). Suppose further that you wish to know whether the universe is in $\sigma$, and you have some computer C to use to answer (all four intelligibility functions of) this question. Then loosely speaking, the prediction complexity of $\sigma$ with respect to C is the minimal amount of Shannon information that must be imposed in C's inputs in order to be assured that C's output correctly answers that question. In particular, if $\sigma$ corresponds to a potential future state of some system S external to C, then $\mathbf{c}(\sigma \mid C)$ is a measure of how difficult it is for C to predict that future state of S.[1s]

In many situations it will be most natural to choose $d\mu$ to be uniform over accessible phase space volume, so that the complexity of *in* is the negative physical entropy of constraining $\hat{u}$ to lie in *in*. But that need not be the case. For example, we can instead define dµ so that the volume of each element of the associated {IN} is some arbitrary positive real number. In this case, the lengths of the elements of {IN} provides us with an arbitrary ordering over those elements.

The following example illustrates the connection between lengths of regions *in* and lengths of strings in TM's:

**Example 3:** In a conventional computer (see Ex. 1 above), we can define a "partial string" s (sometimes called a "file") taking up the beginning of an input section as the set of all "complete strings" taking up the entire input section whose beginning is s. We can then identify the input to the computer as such a partial string in its input section. (Typically, there would be a special fixed-size "length of partial string" region even earlier, at the very beginning of the input section, telling the computer how much of the complete string to read to get that partial string.) If we append certain bits to s to get a new longer input partial string, s', the set of complete strings consistent with s' is a proper subset of the set of complete strings consistent with s. Assuming our measure dµ is independent of the contents of the "length of partial string" region, this means that **l**(s') ≥ **l**(s).

This is in accord with the usual definition of the length of a string used in Turing machine theory. Indeed, if s' contains n more bits than does s, then there are $2^n$ times as many complete strings consistent with s as there are consistent with s'. Accordingly, if we take logarithms to have base 2, **l**(s') = **l**(s) + n.

Say we want our computer to be able to predict whether $\hat{u}$ lies in some set σ. (To maintain the analogy with Turing machines, σ could delineate an "output partial string". This could be done for example by delineating a particular $OUT_p$ value, perhaps even one in some other computer.) In the usual way, this corresponds to having the binary partition { $\hat{u} \in \sigma$, $\hat{u} \notin \sigma$} be weakly predictable to our computer. So the prediction complexity of that prediction is the length of the shortest region of our input space that will weakly induce that prediction. (Note that since we require that all four intelligibility functions of σ be induced, more than one input "partial string" is required for that induction, in general.)

The fact that $OUT_p$ values specify the set $A(OUT_q)$ makes working with Def.'s 10 and 11 a bit messy. In particular, to relate prediction complexity to properties of the associated universal physical computer we must use a set of "identity" intelligibility functions defined as follows:

**Definition 12 (i):** Given a space $X \subseteq \mathbf{B}$ and a physical computer C with input and output spaces {IN} and {OUT} respectively,

$\{I^C_X\}$ is the set of all question-independent intelligibility functions of C where $A(I^C_X) = X$, and where $\forall \hat{u}$ such that $A(OUT_q(\hat{u})) = X$, $I^C_X(\hat{u}) = OUT_\alpha(\hat{u})$.

We also will need the following definition:

**Definition 12 (ii):** Given a space $X \subseteq \mathbf{B}$ and a physical computer C with input and output spaces {IN} and {OUT} respectively, when X is a set $C^{-1}(X)$ is also a set, defined as those $IN \in \{IN\}$ such that $IN(\hat{u}) = IN \Rightarrow A(OUT_q(\hat{u})) = X$.

So for example, if $X = \mathbf{B}$, a pair $(IN^2 \in [C^2]^{-1}(X), I^2_X \in \{I^2_X\})$ is an input to $C^2$ and an intelligibility function of $C^2$'s output, respectively. That input $IN^2$ induces an associated output question, $q^2 \in OUT^2_q$, that takes on (both) $\mathbf{B}$ values as one varies over the $\hat{u}$ input to it. Similarly, the intelligibility function $IN^2_X$ takes on (both) $\mathbf{B}$ values as one varies over the inputs to it.

Using these definitions, we now bound how much more complex a partition can appear to $C^1$ than to $C^2$ if $C^1$ can strongly predict $C^2$. Though somewhat forbidding in appearance, intuitively, the bound simply reflects the complexity cost of "encoding" $C^2$ in $C^1$'s input.

**Theorem 8:** Given any partition $\pi$ and physical computers $C^1$ and $C^2$ where $C^1 \gg C^2 > \pi$,

i) $c(\pi | C^1) - c(\pi | C^2) \leq$

$\ln[o(2^\pi)] - \ln[3] +$

$\max_{\{X \subseteq \mathbf{B},\, IN^2 \in [C^2]^{-1}(X),\, I^2_X \in \{I^2_X\}\}} l[(C^1)^{-1}(C^2, IN^2, I^2_X)] -$

$\min_{\{X \subseteq \mathbf{B},\, IN^2 \in [C^2]^{-1}(X)\}} l[IN^2]$,

or alternatively,

ii) $c(\pi | C^1) - c(\pi | C^2) \leq$

$\ln[o(2^\pi)] +$

$\min_{\{X \subseteq \mathbf{B},\, IN^2 \in [C^2]^{-1}(X),\, I^2_X \in \{I^2_X\}\}} l[(C^1)^{-1}(C^2, IN^2, I^2_X)] -$

$\min_{\{X \subseteq \mathbf{B},\, IN^2 \in [C^2]^{-1}(X)\}} l[IN^2]$ .

**Proof:** Given any intelligibility function f for $\pi$, consider any $IN^2_f \in \{IN^2\}$ that weakly induces f, i.e., such that $IN^2(\hat{u}) = IN^2_f \Rightarrow OUT^2_p(\hat{u}) = (A(f), f(\hat{u}))$. (The analysis will not be affected if $\pi$ is an output partition and we restrict attention to those intelligibility functions for $\pi$ that are question-independent.) Since $C^1 >> C^2$, we can then choose an $IN^1$, $IN^1_f(IN^2_f)$, to strongly induce $IN^2_f$ together with any question-independent intelligibility function of $OUT^2_p$. (Indeed, in general there can be more than one such value of $IN^1$ that induces $IN^2_f$.) So in particular, we can choose it so that the vector $OUT^1_p(\hat{u}) = (A(I^2_{A(f)}), I^2_{A(f)}(\hat{u}))$ for any possible function $I^2_{A(f)}$. Now for that $IN^1$, $IN^2(\hat{u}) = IN^2_f$, and therefore $A(OUT^2_q(\hat{u})) = A(f)$, which means that $I^2_{A(f)}(\hat{u}) = OUT^2_\alpha(\hat{u})$, which in turn equals $f(\hat{u})$ for that $IN^2$. So $\forall \hat{u}$ such that $IN^1(\hat{u}) = IN^1_f(IN^2_f)$, $OUT^1_p(\hat{u}) = (A(f), f(\hat{u}))$. In other words, $IN^1_f(IN^2_f)$ weakly induces in $C^1$ the same intelligibility function for $\pi$ that $IN^2_f$ weakly induces in $C^2$. However since $IN^1(\hat{u}) = IN^1_f(IN^2_f) \Rightarrow IN^2_f(\hat{u}) = IN^2_f$, the set of $\hat{u} \in \hat{U}$ such that $IN^1(\hat{u}) = IN^1_f(IN^2_f)$ is $\subseteq$ the set such that $IN^2(\hat{u}) = IN^2_f$. This means that $l(IN^1_f(IN^2_f)) \geq l(IN^2_f)$. (Our task, loosely speaking, is to bound this difference in lengths, and then to extend the analysis to simultaneously consider all such question-independent intelligibility functions f.)

Take $\{f_i\}$ to be the set of all intelligibility functions for $\pi$. By the preceding construction, $\pi$ is weakly predictable to $C^1$ with a (not necessarily proper) subset of $\{IN^1_{f_i}(IN^2_{f_i})\}$ being a member of $(C^1)^{-1}(\pi)$. Now any member of $(C^1)^{-1}(\pi)$ must contain at least three disjoint elements, corresponding to intelligibility functions q with $A(OUT^1_q(\hat{u})) = \mathbf{B}, \{0\}$, or $\{1\}$. (See the discussion just before Lemma 1.) Accordingly, the volume (as measured by $d\mu$) of any subset of $\{IN^1_{f_i}(IN^2_{f_i})\} \in (C^1)^{-1}(\pi)$ must be at least 3 times the volume of the element of $\{IN^1_{f_i}(IN^2_{f_i})\}$ having the smallest volume. In other words, the length of any subset of $\{IN^1_{f_i}(IN^2_{f_i})\} \in (C^1)^{-1}(\pi)$ must be at most $-\ln(3)$ plus the length of the longest element of $\{IN^1_{f_i}(IN^2_{f_i})\}$. Therefore $\mathbf{c}(\pi \mid C^1) \leq \max_{f_i} [\mathbf{l}(IN^1_{f_i}(IN^2_{f_i}))] - \ln(3)$.

Now take $\{IN^2_{f_i}\}$ to be the set in $(C^2)^{-1}(\pi)$ with minimal length. $\{IN^2_{f_i}\}$ has at most $o(2^\pi)$ disjoint elements, one for each intelligibility function for $\pi$. Using the relation $\min_i[g_i] = -\max_i[-g_i]$, this means that $\mathbf{c}(\pi \mid C^2) \geq -\ln[o(2^\pi)] + \min_{f_i}[\mathbf{l}(IN^2_{f_i})]$. Therefore we can write $\mathbf{c}(\pi \mid C^1) - \mathbf{c}(\pi \mid C^2) \leq \ln[o(2^\pi)] - \ln(3) + \max_{f_i}[\mathbf{l}(IN^1_{f_i}(IN^2_{f_i}))] - \min_{f_i}[\mathbf{l}(IN^2_{f_i})]$. The fact that for all $IN^2_{f_i}$, $IN^2(\hat{u}) = IN^2_{f_i} \Rightarrow A(OUT^2_q(\hat{u})) = A(f_i) \subseteq \mathbf{B}$ completes the proof of (i).

To prove (ii), note that we can always construct one of the sets in $(C^1)^{-1}(\pi)$ by starting with the set consisting of the element of $\{IN^1_{f_i}(IN^2_{f_i})\}$ having the shortest length, and then successively adding other $IN^1$ values to that set, until we get a full (weak) prediction set. Therefore $\mathbf{c}(\pi \mid C^1) \leq \min_{f_i} \mathbf{l}(IN^1_{f_i}(IN^2_{f_i}))$. Using this bound rather than the one involving $-\ln(3)$ establishes (ii). **QED.**

Note that the set of $X \in \mathbf{B}$ such that $[C^2]^{-1}(X)$ exists must be non-empty, since $C^2 > \pi$. Similarly, $C^2 > \pi$ means that there is a $\hat{u}$ such that $A(OUT_q(\hat{u})) = X \subseteq \mathbf{B}$. The associated $I^2_X$ always exists by construction: simply define $I^2_X(\hat{u}) = OUT^2_\alpha(\hat{u}) \forall \hat{u}$ such that $A(OUT_q(\hat{u})) = X$, and for all other $\hat{u}$, $I^2_X(\hat{u}) = x$ for some $x \in X$. Therefore the extrema in our bounds are always well-defined.

As one varies $\pi$, in both bounds in Thm. 8 the dependence of the bound on $C^1$ and $C^2$ does not change. In addition, those bounds are independent of $\pi$ for all $\pi$ sharing the same cardinality. So in particular they are independent of $\pi$ for all binary partitions like those discussed in Ex. 3. This illustrates how Thm. 7 is the physical computation analogue of the result in Turing machine theory that the difference in algorithmic complexity of a fixed string with respect to two separate Tur-ing machines is bounded by the complexity of "emulating" the one Turing machine on the other, independent of the fixed string in question.

Consider the possibility that for the laws of physics in our universe, there exist partitions IN(.) and OUT(.) that constitute a universal physical computer $C^*$ for all other physical computers in our universe. Then by Thm. 5, no other computer is similarly universal. Therefore there exists a unique prediction complexity measure that is applicable to all physical computers in our universe, namely complexity with respect to $C^*$. (This contrasts with the case of algorithmic information complexity, where there is an arbitrariness in the choice of the universal TM used.) If instead there is no universal physical computer in our universe, then every physical computer C must fail at least once at (strongly) predicting some other physical computer. (Note that unlike the case with weak predictability considered in Thm. 2, here we aren't requiring that the universe be capable of having two distinguishable versions of C.) This establishes the following:

**Theorem 9:** Either infallible strong prediction is impossible in our universe, or there is a unique complexity measure in our universe.

Similar conclusions hold if one restricts attention to a set of (physically localized) conventional physical computers (cf. Ex. 1 above), where the light cones in the set are arranged to allow the requisite information to reach the putative universal physical computer.

**FUTURE WORK AND DISCUSSION**

Any results concerning physical computation should, at a minimum, apply to the computer lying on a scientist's desk. However that computer is governed by the mathematics of deterministic finite automata, not that of Turing machines. In particular, the impossibility results concerning Turing machines rely on infinite structures that do not exist in any computer on a scientist's desk.

Accordingly, there is a discrepancy between the domain of those results and that of any truly general theory of physical computers.

On the other hand, when one carefully analyzes actual computers that perform calculations concerning the physical world, one uncovers a mathematical structure governing those computers that is replete with its own impossibility results. While much of that structure parallels Turing machine theory, much of it has no direct analogue in that theory. For example, this new structure has no need for tapes, moveable heads, internal states, read/write capabilities, and the like, none of which have any obvious connection to the laws governing our universe (i.e., any connection to quantum mechanics and general relativity).

In fact, when the underlying functions of real-world computers are stripped down to their essentials, one does not even need to identify a "computer" with a device occupying a particular localized region of space-time, never mind one with heads and the like. In place of all those concepts one has a structure involving several partitions over the space of all worldlines of the universe. The partitions in that structure delineate a particular computer's inputs, the questions it addresses, and its outputs. The impossibility results of physical computation concern the relation of those partitions. Computers in the conventional, space-time localized sense (the box on your desk) are simply special examples, with lots of extra restrictions that turn out to be unnecessary in the underlying mathematics. Accordingly, the general definition of a "physical computer" has no such restrictions. A side-benefit of this breadth is that the associated mathematics can be viewed as concerning many information-processing activities (e.g., observation, control) normally viewed as distinct from computation.

In the first paper in this pair, this definition of a physical computer was motivated and presented, along with some associated theorems. Those theorems imply, amongst other things, that fool-proof prediction of the future is impossible — there are always some questions concerning the future that cannot even be posed to a computer, and of those that can be posed, there are always some for which the computer's answer will be wrong. By exploiting the breadth of the definition of physical "computation", similar results hold for the information-processing of observation and of control. All of this is true even in a classical, non-chaotic, finite universe, and regardless of the where in the Chomsky hierarchy the computer lies.

This second paper launches from the theorems of the first paper into a broader, albeit preliminary investigation of the mathematics of physical computation. It is shown that the computability structure relating distinct physical computers is that of a directed, acyclic graph. In addition, there is at most one computer (called a "god computer") that can predict /observe /control all other computers. Other results derived include limits on error-correction using multiple computers, and some analogues of the Halting theorem.

Next a definition of the complexity of a particular computational task for a particular physical computer, prediction complexity, is motivated. The motivation of this new definition of complexity proceeds by analogy to the concept of the algorithmic information complexity of a symbol sequence for a universal Turing machine. However whereas algorithmic information complexity concerns a Turing machine's generating such a symbol sequence, prediction complexity involves a physical computer's addressing a computational task concerning the physical universe.

The difference in prediction complexity of a particular task $\pi$ for two different physical computers $C^1$ and $C^2$ is considered. It is proven that that complexity difference is bounded by a function that only depends on $C^1$ and $C^2$, and is independent of $\pi$. This bound relating the difference in complexity for two physical computers is analogous to the algorithmic information complexity cost of emulating one universal Turing machine with another one. Finally, it is proven that either a certain kind of computation is not possible in our universe, or there is a preferred computer in our universe. If it exists, that computer could be used to uniquely specify the prediction complexity of any task $\pi$. Accordingly, either a certain kind of computation is impossible, or there is a preferred definition of physical complexity (in contrast to the arbitrariness inherent in algorithmic information complexity's choice of universal Turing machine).

The following ideas are just a few of the questions that the analysis of this paper raises:

i) What other restrictions are there on the predictability relations within distinguishable sets of physical computers beyond that they form unions of DAG's? In other words, which unions of DAG's can be manifested as the predictability relations within a distinguishable set? How does this answer change depending on whether we are considering sets of fully input-distinguishable computers or sets of pairwise-distinguishable computers? For what computers are there finite /

countably infinite / uncountably infinite numbers of levels below it in the DAG to which it belongs? Might such levels be gainfully compared to the conventional computer science theory issue of position in the Chomsky hierarchy?

ii) One might try to characterize the unpredictability-of-the-future result of paper I as the physical computation analogue of the following issue in Turing machine theory. Can one construct a Turing machine M that can take as input A, an encoding of a Turing machine and its tape, and for any such A compute what state A's Turing machine will be in after will be in after n steps, and perform this computation in fewer than n steps? This characterization suggests investigating the formal parallels (if any) between the results of these papers and the "speed-up" theorems of computer science.

iii) More speculatively, the close formal connection between the results of this second paper and those of computer science theory suggest that it may be possible to find physical analogues of most of the other results of computer science theory, and thereby construct a full-blown "physical computer science theory". In particular, it may be possible to build a hierarchy of physical computing power, in analogy to the Chomsky hierarchy. In this way we could translate computer science theory into physics, and thereby render it physically meaningful.

We might be able to do at least some of this even without relying on the DAG relationship among the physical computers in a particular set. As an example, we could consider a system that *can* correctly predict the future state of the universe from any current state of the universe, before that future state occurs. The behavior of such a system is perfectly well-defined, since the laws of physics are fully deterministic (for quantum mechanics this statement implicitly presumes that one views those laws as regarding the evolution of the wave function rather than of observables determined by non-unitary transformations of that wave function). Nonetheless, by the central unpredictability result of paper I, we know that such a system lies too high in the hierarchy to exist in more than one copy in our physical universe.

With such a system identified with an oracle of computer science theory we have the definition of a "physical" oracle. Can we construct further analogues with computer science theory by leveraging that definition of a physical oracle? In other words, can we take the relationships between (computer science) oracles, Turing machines, and the other members of the (computer science) Chomsky hierarchy, and use those relationships together with our (physical) oracle and physical computers to gainfully define other members of a (physical) Chomsky hierarchy?

iv) Can we then go further and define physical analogues of concepts like P vs. NP, and the like? Might the halting probability constant $\Omega$ of algorithmic information theory have an analogue in physical computation theory?

As another example of possible links between conventional computer science theory and that of physical computers, is there a physical computer analogue of Berry's paradox? Weakly predicting a partition is the physical computation analogue of "generating a symbol sequence" in algorithmic information complexity. The core of Berry's paradox is that there are numbers k such that no Turing machine can generate a sequence having algorithmic information complexity k (with respect to some pre-specified universal Turing machine U). So for example one closely related issue in physical computation is to characterize the physical computers $C^1$ and $x \in \mathfrak{R}$ such that $\exists$ a computer $C^2$ where $C^1 \gg C^2$ and where $\forall$ partitions $\pi$, $C^2$ weakly predicts whether $\mathbf{c}(\pi \mid C^1) > x$ (i.e., such that $\exists\ IN^2 \in \{IN^2\}$ such that $IN^2(\hat{u}) = IN^2 \Rightarrow OUT^2_p(\hat{u}) = (\mathbf{B}$, whether $\mathbf{c}(\pi \mid C^1) > x))$.

v) Concerns of computer science theory, and in particular of the theory of Turing machines, have recently been incorporated into a good deal of work on the foundations of physics [33]. Future work involves replacing physical computers for Turing machines in this work, along with replacing notions like prediction complexity for notions like algorithmic complexity.

vi) Other future work involves investigating other possible definitions of complexity for physical

computation. Even sticking to analogues of algorithmic information complexity, these might extend significantly beyond the modifications to the definition of prediction complexity discussed in the text. For example, one might try to define the analogue of a bit sequence's "length" in terms of the number of elements in Q. One might also take the (inverse) complexity of a computational device to be the number of input-distinguishable computers that can predict that device (working in some pre-specified input-distinguishable set, presumably).

vii) Yet other future work includes calculating physical complexity of various systems for some of the simple physical models of real-world computers (e.g., "billiard ball" computers, DNA computing, etc.) that have been investigated, and investigating the prediction complexity of systems like crystals and gases.

**FOOTNOTES**

[1] Especially for non-binary $\pi$, many other definitions of prediction complexity besides Def. 11(ii) can be motivated. For example, one could reasonably define the complexity of $\pi$ to be the sum of the complexities of each binary partition induced by an element of $\pi$, i.e., one could define it as $\Sigma_{p \in \pi} \mathbf{c}(\{\hat{u} \in p, \hat{u} \notin p\} \mid C)$. Another variant, one that would differ from the one considered in the text even for binary partitions, is $\min_{\rho \in C^{-1}(\pi)} [\Sigma_{IN \in \rho} \mathbf{l}(IN)]$. For reasons of space, no such alternatives will be considered in this paper.

**ACKNOWLEDGMENTS:** This work was done under the auspices of the Department of Energy, the Santa Fe Institute, and the National Aeronautics and Space Administration. I would like to thank Bill Macready, Cris Moore, Paul Stolorz, Tom Kepler and Carleton Caves for interesting discussion.